# VIRAL SELF-ASSEMBLY AS A THERMODYNAMIC PROCESS

Robijn F. BRUINSMA[1†], William M. GELBART[2], David REGUERA[2], Joseph RUDNICK[1] and Roya ZANDI[2]

[1]Department of Physics and Astronomy, and [2]Department of Chemistry and Biochemistry, The University of California at Los Angeles, Los Angeles 90095-1569, California, USA

[†]Instituut-Lorentz/LION, Universiteit Leiden, 2300 RA, Leiden, The Netherlands,

ABSTRACT

The protein shells, or capsids, of all sphere-like viruses adopt *icosahedral symmetry*. In the present paper we propose a statistical thermodynamic model for viral self-assembly. We find that icosahedral symmetry is *not* expected for viral capsids constructed from structurally identical protein subunits and that this symmetry requires (at least) two internal "switching" configurations of the protein. Our results indicate that icosahedral symmetry is not a generic consequence of free energy minimization but requires optimization of internal structural parameters of the capsid proteins.

PACS #: 87.15.Nn, 81.16.Dn, 61.50.Ah, 87.16.Dg

Spontaneous self-assembly of simple units into larger structures plays an important role in molecular biology and materials science. A striking example is the self-assembly of *viruses*[1]. As long ago as 1955 Fraenkel-Conrat and Williams showed[2] that an infectious *rod*-like virus – the Tobacco Mosaic Virus (TMV) – could be reversibly reconstituted in the laboratory from a two-component solution of its purified genome (RNA in this instance) and the protein that comprises its cylindrical capsid. Reversible self-assembly has been demonstrated as well for a number of *sphere*-like plant viruses[3]. In all of these cases the assembly proceeds spontaneously, without involving " fuel consumption" such as ATP hydrolysis.

As noted by Crick and Watson[4] (CW) the capsids of viruses are formed from a minimum number of gene products, given the small size of viral genomes. On this basis, CW argued that spherical viruses should actually be in the form of regular polyhedra ("platonic solids") all of whose faces are identical perfect polygons in which all protein units sit in identical environments; the largest shell of this kind is an icosahedron consisting of 60 equivalent sub-units. Subsequent capsid structure determinations confirmed the special role of icosahedral symmetry, but also indicated that *larger* numbers of protein subunits were involved.

Caspar and Klug[5] (CK) proposed a geometrical scheme for the general construction of icoshahedral shells with an arbitrarily large number of subunits. Capsid proteins usually can be grouped into "capsomers" of either hexamer/pentamer units or trimer units. The number of proteins constituting a closed isometric surface equals 60 times a "triangulation" (T) number that adopts special integer values[6] such as 1, 3, 4, and 7 (see Fig.1). Electron and X-ray diffraction studies have confirmed that the T-number classification applies to almost all sphere-like viruses[7].

The success of the CK construction for a broad range of sphere-like viruses indicates that the *production* of icosahedral symmetry might be a generic feature of the capsid free energy. Continuum elasticity theory supports this notion, for large capsids: the deformation energy cost incurred upon closing a hexagonal sheet on itself is minimized when the twelve five-fold sites are located as far as possible from each other, i.e. if the shell adopts icosahedral symmetry[8]. The CK construction is reproduced also in simple models for viral capsids based on the covering of a sphere with disks *provided* icosahedral symmetry is imposed[9]. Icosahedral symmetry is, however, far from obligatory. *In vitro* self-assembly can produce not only icosahedral capsids, but also hexagonal sheets, rod-like aggregates resembling "buckytubes", non-icosahedral sphere-like capsids, and still more complex structures[3,10]. Closed, cone-like structures of hexamers and pentamers, for example, are reported for the lentiviruses, such as HIV[11].

In this Letter we propose a simple free energy for capsid self-assembly that can be used to study under what conditions self-assembly leads to structures with icosahedral symmetry. Our phenomenological Hamiltonian separates the free energy cost of a protein shell into an "in-plane" part describing deformations away from the ideal hexagonal packing structure, and an "out-of-plane" part arising from the difference between the preferred and the actual angle of neighboring capsomers. We computed a self-assembly phase diagram as a function of concentration and "spontaneous curvature". Earlier accounts[12] of capsid self-assembly assume a particular capsid size and structure and focus instead on the *kinetics* of formation of intermediate and final structures. Additionally, a deterministic "local rules" theory has been developed[13] in which the nearest-neighbor interactions between individual proteins provide an intricate, coded template for specific capsid arrangements.

The model in its simplest form treats both hexameric and pentameric capsomers as disks with an adhesive edge that describes the inter-subunit bonding. Capsid self-assembly takes place from a disk solution with a given total mole fraction Φ and chemical potential μ. Let V(θ) be the gain in energy when two disk edges are joined, where θ is the angle between the disk normals. We will assume that

$$V(\theta) = V(0) + \frac{1}{2}\kappa(\theta - \theta^*)^2 . \qquad (1)$$

The V(0) term in Eq. (1) is the (negative) disk-disk *adhesion* energy[14], which is assumed to be the dominant energy scale. The energy κ in the second term corresponds to the *bending stiffness* of a joint, while θ* is the optimal angle of a joint; θ* plays the role of the spontaneous curvature of the capsomer shell[15]. As first proposed by CK[5], spontaneous curvature is the natural thermodynamic control parameter for capsid size.

The disk adhesion energy $V(0)$ favors packing a maximum number of disks on the capsid surface. Let $\rho(N)$ (<1) denote the fraction of the capsid area covered by N disks at their maximum packing density on the surface of a sphere. An upper limit for $\rho(N)$ is the coverage $\rho_{max} = \frac{\pi}{2\sqrt{3}}$ of a flat, hexagonal sheet of disks. By curving a hexagonal sheet to cover a sphere, additional interstitial spaces are introduced in the packing structure (see Fig.1). The loss of binding energy suffered through the introduction of these holes -- the in-plane deformational energy -- is included as a "mean field" term $N(\rho_{max} - \rho(N))^2$, obtained by treating the layer of disks as a stretchable elastic sheet[16]. The resulting capsid Hamiltonian is:

$$H(N) = \frac{z}{2} N V(0) + \frac{B}{2} N(\rho_{max} - \rho(N))^2 + \frac{\kappa}{2} \sum_{i,j} (\theta_{i,j} - \theta^*)^2 . \qquad (2)$$

Here z is the mean number of nearest neighbors per disk and B is a two-dimensional compressional modulus (times the disk area). The sum in the last term runs over all nearest-neighbor pairs of disks, with $\theta_{i,j}$ the angle between their normals.

The problem of finding the coverage $\rho(N)$, of a sphere that is optimally (close-) packed by N circular disks is known in the mathematical literature as the *Tammes Problem*[17]. Results are available for small N values either in exact or numerical form[18]. Figure 2a shows $\rho(N)$ for N ranging from 10 to 75. Note that $\rho(N)$ remains significantly below the asymptotic value $\rho_{max}$, even for the largest available N values. The capsid energy $H(N)$, as computed from Eq. (2), exhibits as a function of N a complex, $\theta^*$-dependent, spectrum of minima (not shown). To construct the self-assembly phase-diagram, we treat N as a statistical quantity with $\Phi(N)$ the mole fraction of N-disk capsids. Minimization of the solution free energy F = < H(N) > - TS with $S = -k_B \sum_N \Phi(N) \ln \Phi(N)$ the mixing entropy, leads to a classical formula of self-assembly[19]: $\Phi(N) \propto \exp^{\beta(\mu N - H(N))}$. The onset of capsid formation is then identified by the condition that half of the disks remain in solution while the other half are incorporated in capsids. Minimization of F leads to the condition $\mu(\Phi)(N^*-1) = H(N^*) - k_B T \ln N^*$, with N* the number of capsomers of the dominant capsid structure at onset.

Figure 2b shows the resulting self-assembly plot in terms of $\Phi$ and $\theta^*$ (for B=10$\kappa$ and $\kappa$=10$k_B$T). For N values up to 75, icosahedral symmetry is in principle possible for N=12 (T=1), N=32 (T=3), N=42 (T=4), and N=72 (T=7), the first four structures in the T-series. The T=1 dodecahedron is indeed quite stable for higher spontaneous curvatures, appearing at very low disk concentrations. However, the capsid structure that appears next for decreasing $\theta^*$ is *not* the T=3 (truncated) icosahedron at N=32 but instead, at N=24, a surprising octahedral, chiral structure that has the symmetry of an archimedean solid known as the *snub-cube* (see the second column in Fig.1). The next structure encountered for decreasing $\theta^*$ is indeed N=32 (though the disk configuration does not have exact icosahedral symmetry[20]), though the N=42 (T=4) structure is superseded by N=48. Finally, N=72 appears, corresponding to T=7 (though again lacking perfect icosahedral symmetry). Note that this series of "magical numbers" coincides with the dominant maxima of $\rho(N)$ (see arrows of Fig.2a). Which of these magical numbers is selected is determined by the spontaneous curvature.

The essential aspect of our result is that, though N=12, N=32, and N=72 do appear as possible capsid geometries, the N=42 (T=4) structure is absent from the spectrum of magical numbers. On the other hand, two completely *non*-icosahedral conformations (N=24 and 48) are predicted to be present. An important test case in this respect is provided by the *polyoma virus*, which is exceptional in that all its capsomers have the same size (with five proteins per capsomer)[21], as assumed in our model. The native form of the polyoma virus is the N=72 (T=7) structure, but self-assembly of polyoma capsid proteins *alone* (i.e., without their genome) produces three dominant structures[22] (depending on pH and ionic strength): N=12, N=24, and N=72. The N=24 structure has the symmetry of a (left-handed) snub-cube, consistent with our model for the case of identical capsomers. We conclude that the adoption of icosahedral symmetry is **not** a generic feature of the self-assembly of finite-sized closed shells constructed from (more than twelve) identical capsomers.

The capsomers of typical viruses are however not all identical as assumed in the simple model. Detailed structure studies[23] show that the twelve pentameric capsomers can have a quite different internal structure, often involving a *conformational switch*[24]. To account for this structural difference, we generalized the model by allowing twelve of the circular disks to have a smaller diameter than the others, and assigning a "switch energy" cost ΔE to each transformation of a protein from a hexameric to a pentameric configuration. The ratio of the small and large disk diameters was chosen to *optimize* the coverage. Recomputing the close-packed structures for this "two-radius" model, we obtained surprising results; the Hamiltonian energies of the N=32 (T=3) icosahedral structure and, in particular, the N=42 (T=4) icosahedral structure now were significantly *lower* than that of the N=24 snub-cube. This was, due to an increase in sphere coverage: ρ(32) increases from 0.846 to 0.89 and ρ(42) from 0.83 to 0.90, which even exceeds ρ(12)! The coverage increase for N=72 (T=7) was more modest, from 0.85 to 0.86. On the other hand, because for the N=24 snub-cube the disks are all in equivalent positions, changing the size of one of the disks only *reduces* the coverage. We conjecture that this virtual elimination of the deformational energy term through the introduction of two, slightly different, disk radii, *only takes place for N=10T+2*, i.e., for the icosahedral symmetry T-structures.

The appearance of a conformational switching energy term of 60ΔE in Eq. (2) leads to an interesting effect: for increasing values of ΔE, *sphere-like capsids transform to rod-like capsids*. This can be understood by comparing the total energy of M spherical capsids, each having a curvature θ equal to the optimal curvature θ*, with that of a single sphero-cylindrical capsid with the same *mean* curvature and the same total area. The last term in Eq. 2 favors a minimum value of the second moment of the curvature distribution. It then follows that the cylinder bending energy is larger than that of the spheres by an amount of order Mκ. On the other hand, each sphere-like capsid must accomodate twelve of the smaller disks, at a total energy price of order 60MΔE. A sphere-to-cylinder transition is thus expected when 60*ΔE/κ* is of order one. We conclude that, with spontaneous curvature, ΔE should be a second important control parameter for capsid size.

Figure 3a shows the self-assembly diagram for the two-radius model as a function of *ΔE/κ* and θ*, giving the structures that appear first for increasing Φ (only the T=1 and T=3 icosahedral structures and infinite tubes were included). We compared this self-assembly diagram with the phase-behavior of a well-studied example of virus reconstitution, namely the T=3 plant virus Cowpea Chlorotic Mottle Virus (CCMV), which is characterized by well-defined pentameric and hexameric capsomers[25]. The equilibrium phase diagram of the capsid proteins (again without genomic material)[26,3] – see Fig. 3b – displays a number of structures: hollow single and multi-shell capsids, hexagonal sheets, and buckytube-like sphero-cylinders[27]. A fairly monodisperse T=3 capsid phase is encountered in the pH range below pH 5.5 while at neutral pH (cross) the dominant population consists of protein dimers for low protein concentrations and buckytubes for higher protein concentrations. Comparing Figs. 3a and 3b it follows that the two-radius model could account for the CCMV phase-diagram if the capsid spontaneous curvature $\theta^*$ increases with salinity, which is reasonable since the CCMV capsid proteins are strongly cationic and if the switching energy $\Delta E$ increases with pH. This last trend actually has been argued to be the case for CCMV proteins (due to titratability of terminal carboxyls) on the basis of structural studies[25].

In summary, we have proposed a simple model Hamiltonian for viral self-assembly by disk-like capsomers. The preferred number of *identical* capsomers is characterized by a sequence of "magical numbers" that does not coincide with the Caspar-Klug sequence. We find that the appearance of icosahedral symmetry for smaller capsids is *not* an automatic consequence of free energy minimization – as it would be in the continuum limit – but instead requires optimization of a structural parameter (the ratio of the two disk radii). The two-radius model reproduces in that case the preference for icosahedral symmetry as well as a sphere-to-rod transition that has been observed for a number of viruses. The structural optimization is presumably the result of some form of biological *adaptation*[22] but this lies beyond the range of the present study.

Acknowledgements: We would like to thank C.Henley, J.Johnson, C.Knobler, D.Nelson and A.McPherson for helpful discussions. On of us (D. R) acknowledges support from the NSF through grant CHE-0076384.

## Figure Captions

Figure 1, First row: the T=1, T=3, T=4 and T=7 capsids produced by the Caspar-Klug construction with 12 pentamers and 10(T-1) hexamers per capsid. T=1 is a dodecahedron and T=3 a truncated icosahedron. Second row: optimal packing arrangements of N disks covering a sphere, known as the Tammes problem. Only N=12 has icosahedral symmetry; N=24 is an archimedean solid known as the snub-cube; N=32 could adopt T=3 icosahedral symmetry but in fact has $D_5$ symmetry. Third row: the improved coverage for N=32, N=42 and N=72 when 12 of the disks adopt a diameter equal to 0.934 of the other disks.

Figure 2a: The sphere coverage for the Tammes problem. The maximum coverage $\frac{\pi}{2\sqrt{3}} = \rho_{max}$, is indicated by the dashed line. The structures encountered in the self-assembly diagram are indicated by arrows. N=24 is the snub-cube. Figure 2b: Self-Assembly Phase Diagram for the single-radius model with B equal to 100 $k_BT$ per disk and κ equal to 10 $k_BT$. The horizontal axis is the spontaneous curvature θ*; the vertical axis is $\frac{k_BT}{\kappa}\ln(\Phi/\Phi_0)$, with Φ the capsomer mole fraction and $\Phi_0 \propto \exp^{1/2(zV(0)/k_BT)}$.

Figure 3: Comparison between the self-assembly phase diagram of the two-radius model (Fig.3a) with that measured for the Cowpea Chlorotic Mottle Virus (a small, T=3, plant virus), Fig.3b. The cross in fig.3b indicates physiological conditions. The horizontal axis in Fig.3a is the pentamer/hexamer switching energy per protein (in units of κ) while the vertical axis is the spontaneous curvature. The dashed line in Figs.3a and b shows the proposed correspondence between ΔE, pH and θ*, salinity. Fig. 3b is adapted from the phase diagram in Ref. 26.

Figure 1:

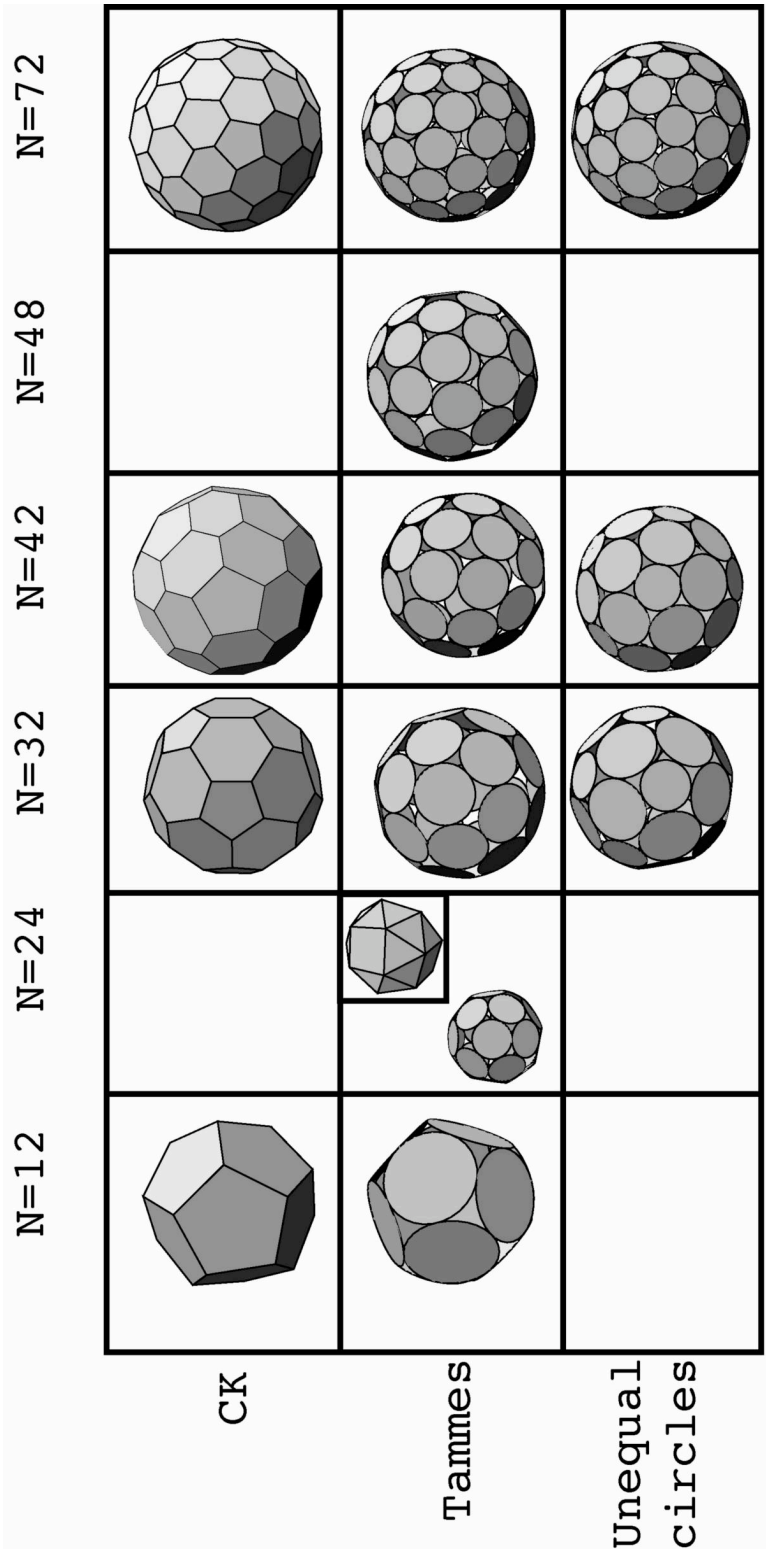

Figure 2:

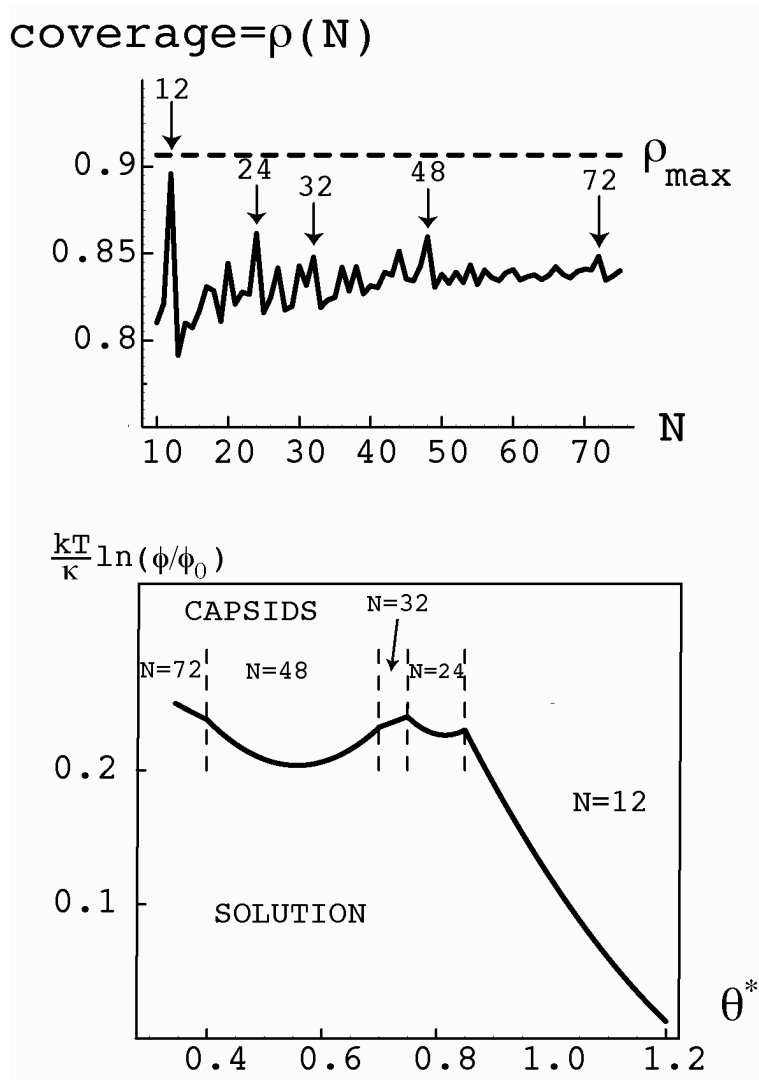

Figure 3:

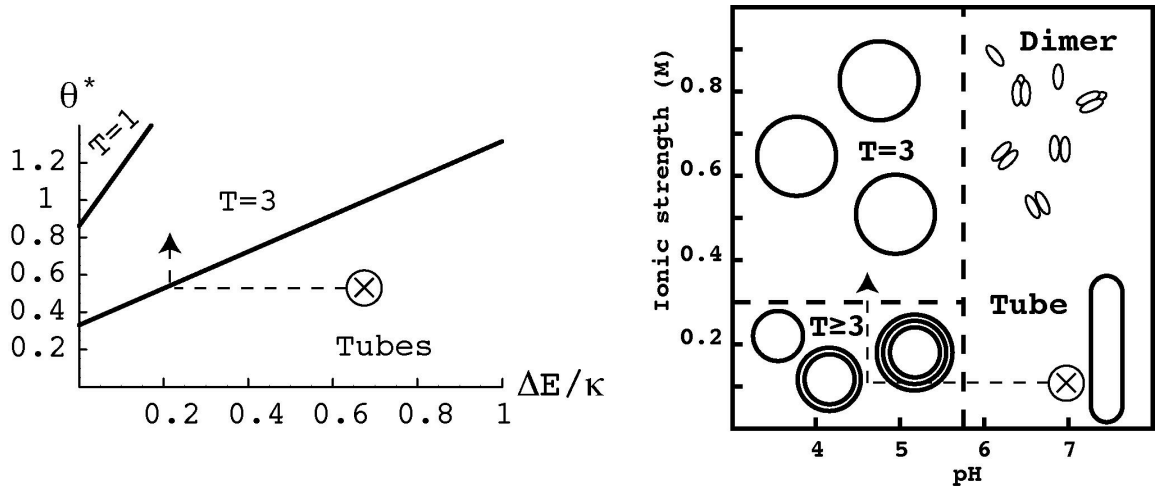